\documentclass[ nofootinbib,amsmath,amssymb,superscriptaddress,aps,prfluids]{revtex4-2}

\usepackage{graphicx}
\usepackage{dcolumn}
\usepackage{bm}
\usepackage{multirow}

\begin{document}

\title{Localised eigenmodes in a moving frame of reference representing convective instability}

\affiliation{
 Aerospace Faculty, Texas A\&M University, United States 
}%
\affiliation{
Aerospace \& Mechanical Engineering, University of Li\`ege, Belgium\\
}%
\affiliation{
Faculty of Aerospace Engineering, Delft University of Technology, The Netherlands\\ 
}%

\author{Koen J. Groot}
\email{koengroot@tamu.edu}
\affiliation{
 Aerospace Faculty, Texas A\&M University, United States 
}%
\affiliation{
Faculty of Aerospace Engineering, Delft University of Technology, The Netherlands\\ 
}%

\author{S\'ebastien E.M. Niessen}
\email{sebastien.niessen@uliege.be}
\affiliation{
Aerospace \& Mechanical Engineering, University of Li\`ege, Belgium\\
}%
\affiliation{
Faculty of Aerospace Engineering, Delft University of Technology, The Netherlands\\ 
}%

\date{\today}

\begin{abstract}
When representing convective instability mechanisms with the streamwise BiGlobal stability approach, results suffer from a sensitivity to the streamwise domain truncation length and boundary conditions. The methodology proposed in this paper resolves this sensitivity by considering a moving frame of reference. In that frame, the spectrum features discrete eigenvalues whose corresponding eigenfunctions decay exponentially in both the up- and downstream directions. Therefore, the truncation boundaries can be placed far enough that both variations in the domain length and artificial boundary conditions have no impact. The discrete nature of the spectrum enables the use of local and non-local stability methods to perform an independent approximation of the BiGlobal eigensolutions via global mode theory. We demonstrate that retrieving set-up-independent solutions in the stationary frame of reference is likely impossible for the considered flow.
\end{abstract}

\maketitle

\section{Introduction}
Linear stability methods are important computational tools for the prediction of the laminar-turbulent transition of boundary layers. The methods deployed by the industry are LST (Linear Stability Theory, \cite{MackAGARD}), which accounts for one-dimensional flow properties only, and PSE (Parabolised Stability Equations, \cite{Herbert1997}), which improves upon LST by marching downstream, allowing to account for small streamwise changes in the flow. The streamwise BiGlobal stability approach is superior to both LST and PSE, in the sense that it can represent all linear perturbation dynamics of practically arbitrary two-dimensional flows \citep{Theofilis2003}. 

Practice has shown, however, that representing convective instability mechanisms, specifically, with the streamwise BiGlobal stability approach suffers from a notorious sensitivity to the computational set-up. This particularly involves the truncation of the domain in the streamwise direction, e.g.\ the specific choice of the domain length and boundary conditions \citep{EhrensteinGallaire,AlizardRobinet,Rodriguez2011}.
The results of \citet[\S IV.B.2]{AlizardRobinet} suggest that the part of the spectrum that is of interest tends to a continuum as the domain length tends to infinity. In regard to this expected limit of the spectrum, \citet{Theofilis2003} states that: `the discretised approximation of the continuous spectrum will always be under-resolved.' The main goal of the proposed methodology in the present paper is to solve the sensitivity issues related to the truncation of the domain. 

Next to the convergence problems due to the domain truncation, the physical interpretation of the BiGlobal stability results, again specifically in the case of convective mechanisms, poses an entirely independent issue. The relationship between global and (non-)local stability approaches is known if a global instability exists, e.g.\ see \cite{monkewitz1993global} and \cite{siconolfi2017towards}. This article provides a first step in developing the physical interpretation of the BiGlobal results in the case of convective instabilities by establishing the link between the BiGlobal stability results on the one hand and $\text{(non-)}$local methods like LST and PSE on the other. \citet{AlizardRobinet} and \citet{Rodriguez2010} have demonstrated the relationship between the solutions corresponding to these different methods by using the complex frequency provided by BiGlobal simulations as the input for the LST and PSE approaches. Using the currently proposed methodology, the link can be made without making use of the BiGlobal stability results. 

The paper is structured as follows: the methodologies used to obtain the BiGlobal and base flow solutions are discussed in \S \ref{sec:methodology}, the results are presented in \S \ref{sec:BiGresults} and the link with LST and PSE solutions is established in \S \ref{sec:linkLSTPSE}. The paper is concluded in \S \ref{sec:conclusion}.

\section{Methodology}
\label{sec:methodology}
It is this article's goal to demonstrate that the sensitivity to the computational set-up can be resolved by formulating the problem in a moving reference frame. An extensive theoretical motivation is presented by \citet[\S 8.6.2]{GrootPhDThesis} and \citet[Appendix A]{groot2019accurate}. 
The streamwise BiGlobal stability problem has been considered in a moving reference frame before \citep{mittal2007stabilized}, but never with the aim of resolving the currently targeted issues. 

\subsection{Streamwise BiGlobal stability problem in a moving reference frame}
In the streamwise BiGlobal stability problem, one considers an infinitesimally small perturbation to a base flow, whose variables are denoted as $\overline{Q}$. Base flows are considered that depend only on the wall-normal $y$- and streamwise $\bar{x}$-coordinate, where the bar denotes the stationary reference frame, in which the base flow is independent of time $\bar{t}$. 

We reformulate the perturbation problem in a reference frame that moves downstream with the constant speed $c_\mathrm{g}$. To this end, the $\bar{x}$- and $\bar{t}$-derivatives in the governing equations have to be transformed as follows:
\begin{equation}
\label{eq:movingreferenceframe}
\frac{\partial }{\partial \bar{x}} = \frac{\partial}{\partial x}, \qquad \frac{\partial }{\partial \bar{t}} = \frac{\partial}{\partial t} - c_\mathrm{g}\frac{\partial}{\partial x}, 
\end{equation}

\noindent where $x = \bar{x} - c_\mathrm{g} \bar{t}$ and $t = \bar{t}$ correspond to the moving reference frame; $y$-derivatives remain unchanged. While independent of $\bar{t}$, the base flow does depend on $t$. Taylor expanding the base flow variables for the elapsed time $\Delta t = t - t_0$ yields:
\begin{eqnarray}
\overline{Q}(\bar{x}(x,t),y)
&= \overline{Q}(\bar{x}(x,t_0),y) + c_\mathrm{g}\Delta t\frac{\partial \overline{Q}}{\partial \bar{x}}(\bar{x}(x,t_0),y) + O\left( \frac{(c_\mathrm{g}\Delta t)^2}{2!}\frac{\partial^2 \overline{Q}}{\partial \bar{x}^2}(\bar{x}(x,t_0),y)\right),
\end{eqnarray}

\noindent where $t_0$ is a reference time. This reveals that the base flow can be assumed to be constant in $t$ when permitting an error of $O\big(c_\mathrm{g} \Delta t\, {\partial \overline{Q}}/{\partial \bar{x}}\big)$, which is small when $c_\mathrm{g}$, $\Delta t$ or $\partial \overline{Q}/\partial \bar{x}$ is small. When $\Delta t = 0$, the solutions are exact. For non-zero $\Delta t$, the solutions of the present approach can be time-integrated such that all unsteady effects due to the moving reference frame are accounted for. Accordingly, the introduced model error can be removed for non-zero $\Delta t$, but this lies out of the current scope. Therefore the time-evolution of the solutions is here discarded.

Upon neglecting the time-dependence of $\overline{Q}(\bar{x}(x,t),y)$, the perturbation problem in the moving reference frame has constant coefficients. In that case, a two-dimensional perturbation variable $q'$ can be represented as the product of an eigenfunction $\tilde{q} = \tilde{q}(x,y)$ and an exponential function of time $t$:
\begin{equation}
\label{eq:ansatz}
q'(x,y,t) = \tilde{q}(x,y)\,\mathrm{e}^{-\mathrm{i}\omega t} + c.c.,
\end{equation}

\noindent where the eigenvalue $\omega$ is a complex angular frequency and $c.c.$ the complex conjugate. The subscripts $r$ and $i$ will denote real and imaginary parts, respectively.

Substituting equations \eqref{eq:movingreferenceframe} and \eqref{eq:ansatz} into the linearised incompressible Navier-Stokes equations yields the streamwise BiGlobal stability equations for a moving reference frame:
\begin{subequations}
\label{eq:BiG}
\begin{align}
\label{eq:BiGx}
-\mathrm{i}\,\omega\,\tilde{u}+\left(\overline{U} - c_\mathrm{g}\right)\frac{\partial \tilde{u}}{\partial x}+\overline{V}\,\frac{\partial\tilde{u}}{\partial y}+\tilde{u}\, \frac{\partial\overline{U}}{\partial x}+\tilde{v}\,\frac{\partial\overline{U}}{\partial y}&=-\frac{\partial \tilde{p}}{\partial x}+\frac{1}{Re}\left(\frac{\partial^2}{\partial x^2} + \frac{\partial^2}{\partial y^2}\right) \tilde{u}; \\
\label{eq:BiGy}
-\mathrm{i}\,\omega\,\tilde{v}+\left(\overline{U} - c_\mathrm{g}\right)\frac{\partial \tilde{v}}{\partial x}+\overline{V}\,\frac{\partial\tilde{v}}{\partial y}+\tilde{u}\, \frac{\partial\overline{V}}{\partial x}+\tilde{v}\,\frac{\partial\overline{V}}{\partial y}
&=-\frac{\partial \tilde{p}}{\partial y}+\frac{1}{Re}\left( \frac{\partial^2}{\partial x^2}+\frac{\partial^2}{\partial y^2}\right) \tilde{v}; \\
\label{eq:BiGc}
\frac{\partial \tilde{u}}{\partial x} + \frac{\partial \tilde{v}}{\partial y}&=0.
\end{align}
\end{subequations}

\noindent The stationary reference frame corresponds to $c_{\mathrm{g}} = 0$. To illustrate how the results in the moving reference frame manifest themselves in the stationary reference frame, define $\bar{\omega}_{\tilde{u}}$ and $\bar{\omega}_{\tilde{v}}$ as follows:
\begin{subequations}
\label{eq:generalisedDoppler}
\begin{align}
\label{eq:generalisedDoppleru}
\omega = \bar{\omega}_{\tilde{u}}(x,y) + \mathrm{i}\,\frac{c_{\mathrm{g}}}{\tilde{u}}\frac{\partial \tilde{u}}{\partial x}(x,y), \qquad \text{where:} \quad \bar{\omega}_{\tilde{u}} = \bar{\omega}_{\tilde{u}}(x,y);\\
\label{eq:generalisedDopplerv}
\omega = \bar{\omega}_{\tilde{v}}(x,y) + \mathrm{i}\,\frac{c_{\mathrm{g}}}{\tilde{v}}\frac{\partial \tilde{v}}{\partial x}(x,y), \qquad \text{where:} \quad\bar{\omega}_{\tilde{v}} = \bar{\omega}_{\tilde{v}}(x,y).
\end{align}
\end{subequations}

\noindent By substituting equation \eqref{eq:generalisedDoppleru} for $\omega$ into equation \eqref{eq:BiGx}, the $c_{\mathrm{g}}$-term is eliminated from equation \eqref{eq:BiGx}. The same can be done with equations \eqref{eq:BiGy} and \eqref{eq:generalisedDopplerv}. This demonstrates that system \eqref{eq:BiG} governs solutions in a stationary reference frame, while permitting a `varying eigenvalue': $\bar{\omega}_{\tilde{u}}$ and $\bar{\omega}_{\tilde{v}}$ are functions of $x$, $y$, $\tilde{u}$ and $\tilde{v}$. The right hand sides of equations \eqref{eq:generalisedDoppler} add up to a constant $\omega$, such that no extra $x$- and $y$-derivatives are introduced when substituting ansatz \eqref{eq:ansatz}. The real part of equations \eqref{eq:generalisedDoppler} corresponds to a Doppler shift, while the imaginary part represents the instantaneous advection-induced $\bar{x}$-translation of the amplitude distribution. 

The only physically substantiated boundary condition for external incompressible flow problems is given at solid walls. There no-slip is imposed for $\tilde{u}$ and $\tilde{v}$ and the Laplace equation for $\tilde{p}$, consistent with the conditions proposed by \citet{theofilis2017linearized}.

\subsection{Base flow}
\label{sec:baseflows}
Stability results, obtained with any stability method, are notoriously sensitive to the base flow, see \cite{Arnal:agard1994}. In addition to this, the solutions of the streamwise BiGlobal problem are also sensitive to the computational set-up. The latter sensitivity is the focal point of this study. Therefore, we consider the self-similar Blasius boundary layer as our base flow. It does not satisfy the full Navier-Stokes equations, but it is slowly-developing, universal and obtainable to arbitrary precision, eliminating all uncertainty associated with the base flow. We would like to emphasise that the goal of this paper is not to analyse the stability of the flow over a flat plate, but rather to demonstrate that the inherent methodological sensitivity is removed by considering the proposed technique. 

The self-similar boundary-layer solution is computed with DEKAF, for all details see \citet{groot2018dekaf}, setting the Mach number $M = 0$. The largest Newton-Raphson residual
corresponds to $\partial^2 \overline{U}/\partial y^2$ and equals $O(10^{-15})$. $N_{\eta} = 500$ nodes were used in the wall-normal direction, yielding at most $O(10^{-12})$ differences with the solution on a grid with $2N_{\eta}-1$ nodes. Both $\overline{U}$ and $\overline{V}$ and their $y$-derivatives are obtained from the self-similar solutions, which are then `GICM-interpolated' onto the BiGlobal grid's collocation nodes. GICM (Groot-Illingworth-Chebyshev-Malik) is an iterative procedure that ensures the spectral accuracy of the Chebyshev discretisation after the inversion of the Malik collocation point mapping and the Illingworth self-similarity transformation. The $x$-derivatives of the base flow quantities are determined with the spectral differentiation matrix corresponding to the BiGlobal problem's discretisation.

\subsection{Numerical set-up}
The solutions to system \eqref{eq:BiG} are obtained numerically on a domain that is truncated in the up- and downstream directions, at $x=x_{\mathrm{in}}$ and $x=x_{\mathrm{out}}$, and far from the flat plate at $y=y_{\mathrm{max}}$  ($x=0$ corresponds to the leading edge and $y=0$ to the wall). At the latter boundary, all perturbation variables are zeroed. The truncation boundaries at $x=x_{\mathrm{in}}$ and $x=x_{\mathrm{out}}$ are respectively referred to as the in- and outflow boundaries. The streamwise domain length is denoted by $L = x_{\mathrm{out}} - x_{\mathrm{in}}$. The literature presents several attempts in prescribing reasonable boundary conditions at the in- and outflow boundaries, see \cite{AlizardRobinet}, \citet[\S 5.4.3]{Rodriguez2010}, \citet[chapter 8]{GrootMScThesis} and \citet{groot2015closing}. Our present aim is to ensure that the solutions are \textit{independent of all truncation boundary conditions} through the use of the moving reference frame. Unless stated otherwise, we use Neumann conditions at the in- and outflow boundaries. This allows revealing when the solutions become dominant at the boundaries and, in turn, when they do become dependent on the boundary conditions.
The problem is discretised with Chebyshev collocation \citep{Canuto1} in both $x$ and $y$. A BiQuadratic mapping \citep{groot2018secondary} is used in the $x$-direction, mapping one-third of the collocation points in-between the points $x_{i1}$ and $x_{i2} > x_{i1}$, each lying within $[x_{\mathrm{in}},x_{\mathrm{out}}]$. The values $x_{i1} = x_{\mathrm{in}} + \frac{1}{3}(x_{\mathrm{out}} - x_{\mathrm{in}})$ and $x_{i2} = x_{\mathrm{in}} + \frac{2}{3}(x_{\mathrm{out}} - x_{\mathrm{in}})$ are used for all presented results. The Malik mapping \citep{malik1990numerical} is used for the wall-normal direction $y$, mapping half the collocation nodes above and below $y_i$. 

Velocity and length scales are respectively made non-dimensional with the freestream speed $\overline{U}\hspace{-.5mm}_e$ and the `global' Blasius length
$\ell = \nu/\overline{U}\hspace{-.5mm}_e$, 
where $\nu$ is the kinematic viscosity. According to this choice, 
$Re = 1$ (using a different scaling had no impact on the numerical results). Table~\ref{tab:refparms} presents the parameters used for the selected reference case. The Arnoldi algorithm is used to solve the discretised problem for the 1000 smallest eigenvalues. 

\begin{table*}[h]
  \begin{center}
\def~{\hphantom{0}}
\begin{ruledtabular}
  \begin{tabular}{c c ccccc c cccc}
   $c_{\mathrm{g}}/\overline{U}\hspace{-.5mm}_e$  & {} {} &   $N_x$  & $x_{\mathrm{in}}/\ell$  &  $x_{i1}/\ell$ & $x_{i2}/\ell$  & $x_{\mathrm{out}}/\ell$ & {} {} & $N_y$  &  $y_{i}/\ell$ & $y_{\mathrm{max}}/\ell$ \\[3pt]
   0.415 & &   300   & $0.2 \times 10^5$ & $2.8 \times 10^5$ & $5.4 \times 10^5$ & $8.0 \times 10^5$ & & 50 & $4.0 \times 10^3$ & $1.6 \times 10^5$\\
  \end{tabular}
\end{ruledtabular}
  \caption{Reference case parameters (not rounded, yielding largest $\omega_i$ for a 3 digit $c_{\mathrm{g}}$-value). }
  \label{tab:refparms}
\vspace{-0mm}
  \end{center}
\end{table*}

\section{Results}
\label{sec:BiGresults}
\begin{figure*}[h]
\includegraphics[width=\textwidth]{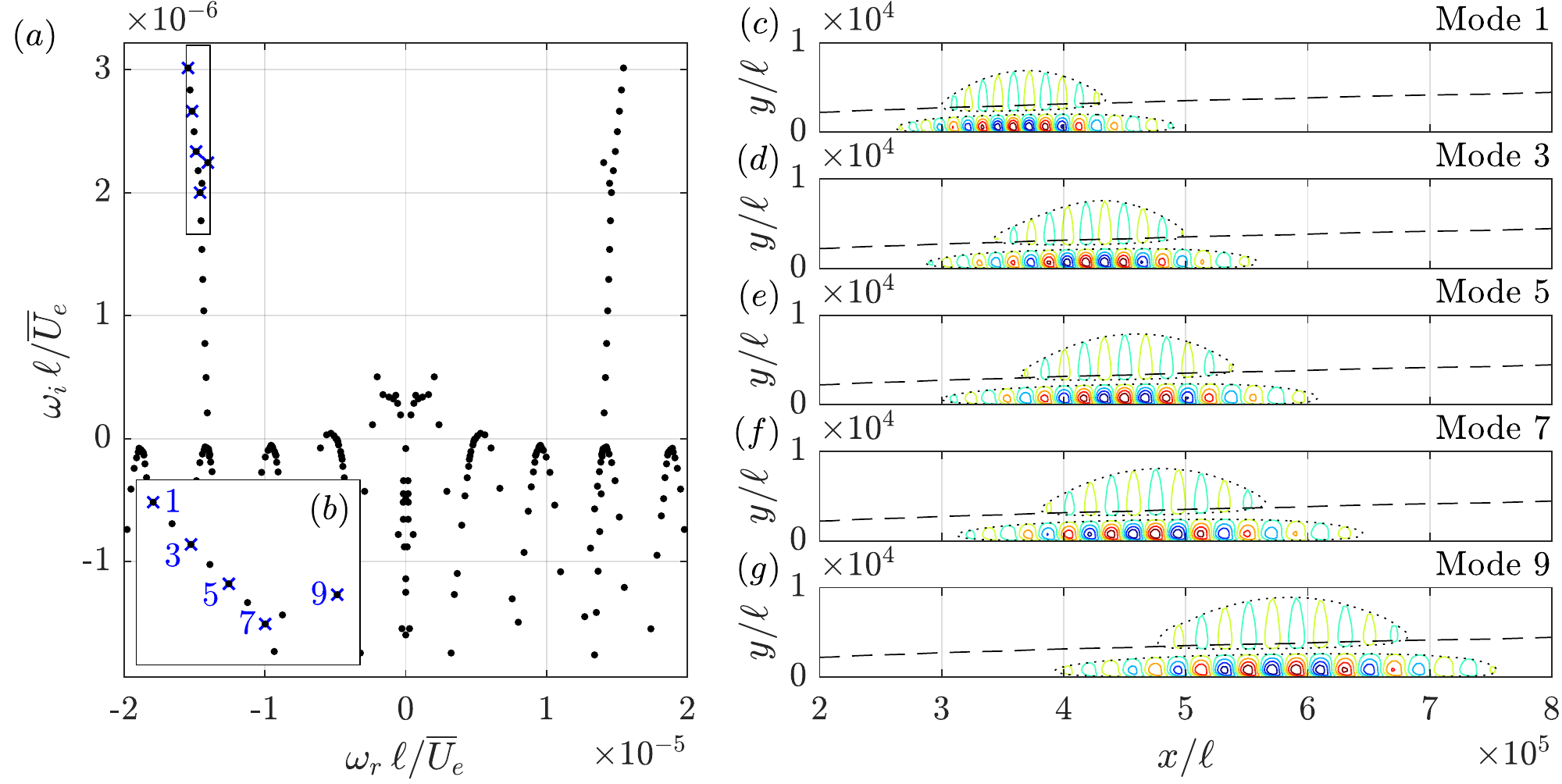}
\vspace{-0mm}
\caption{($a$) $\omega$-spectrum, ($b$) zoom on box and ($c$--$g$) isocontours of the real part of $\tilde{u}$ (coloured lines, from min- (blue) to maximum (red) with $\Delta = 2/9$, scaling the maximum to unity) and $|\tilde{u}|$ (dotted, level: 1/9) corresponding to the eigenvalues labelled in ($b$) for the reference case $c_{\mathrm{g}}/\overline{U}\hspace{-.5mm}_e = 0.415$, $\delta_{99}$-isocontour (dashed).}
\label{fig:existentialgraph}
\vspace{-0mm}
\end{figure*}

The spectrum and $\tilde{u}$-eigenfunctions of interest for the reference case $c_{\mathrm{g}}/\overline{U}\hspace{-.5mm}_e = 0.415$ are shown in Fig.~\ref{fig:existentialgraph}. The attention is restricted to eigenvalues with negative $\omega_r$-values (marked blue in Fig.~\ref{fig:existentialgraph}($b$)). The reference case and eigenmode selection will be justified in \S \ref{sec:variationcg}. The modes of interest form a branch with 3 sub-branches: the top-left `main' branch, housing modes labelled 1 to 5, the rightward `side' branch, in which mode 9 resides, and the downward branch, accommodating mode 7, that appears to continue indefinitely into the stable half-plane. The selected eigenmodes represent wave packets: all eigenfunctions ($\tilde{u}$, $\tilde{v}$ and $\tilde{p}$) decay exponentially toward all truncation boundaries. The instantaneous propagation speed of these wave packets is equal to the speed of the reference frame, therefore this justifies referring to the latter speed as a group speed. The solutions along the main branch are independent of the numerical set-up, this is demonstrated next. 

\subsection{Independence of the numerical set-up and truncation boundary conditions}
\label{sec:BiG:numpars}
Convergence information is given in Table~\ref{tab:conv} for several of the labelled modes in Figs.~\ref{fig:existentialgraph}($a$) and \ref{fig:cgsweepgraphreal}($a$). The relative error in the eigenvalue's magnitude is determined by varying the following numerical aspects independently: the streamwise domain length (indicated by $\epsilon_L$, fixing the relative resolution $N_x/L$), the resolution in the streamwise direction ($\epsilon_{N_x}$), the boundary conditions ($\epsilon_{\mathrm{BC}}$) and the domain height ($\epsilon_{y_{\mathrm{max}}}$, fixing the resolution in the boundary layer by keeping $N_y$ and $y_i$ constant). Overall, relative errors of $O(10^{-4})$ are attained. When representing convective instability mechanisms with the streamwise BiGlobal approach, these small errors are unprecedented in the sense that spectra computed in the stationary frame of reference presented in literature experienced $O(1)$ errors while changing the streamwise domain length. Before elaborating further, it should be noted that $\epsilon_{y_{\mathrm{max}}}$ is the largest contributor to the overall eigenvalue error. Now that the issues related to the streamwise direction are tackled, $\epsilon_{y_{\mathrm{max}}}$ features the slowest convergence rate. Accordingly, the selection of the reference case and the convergence study were approached by reducing $\epsilon_{y_{\mathrm{max}}}$ to a reasonably low level and using that level as an upper bound for the other errors.

\begin{table*}[h]
  \begin{center}
\vspace{-0mm}
\def~{\hphantom{0}}
\begin{ruledtabular}
  \begin{tabular}{r c c c c}
\multirow{4}{*}{\hspace{-3mm}$\begin{array}{r}
\text{Mode} \\ \text{properties}
\end{array}\left| \begin{array}{c}
{}\\{}\\{}\\{}\\{}
\end{array} \right.$ \hspace{-5mm}} & Mode$\,$\#  & {}\;\hspace{15mm}1 & {}\;\hspace{5mm}5 & {}\;\hspace{10mm}10 \\[3pt]
 & $c_{\mathrm{g}}/\overline{U}\hspace{-.5mm}_e$  & {}\;\hspace{14mm}0.415 & {}\;\hspace{4mm}0.415 & {}\;\hspace{9mm}0.470 \\[3pt]
&  $\omega_r\,\ell/\overline{U}\hspace{-.5mm}_e$ & $-1.546\underline{07445982}
  \times 10^{-5}$ & $-1.488\underline{26}
  \times 10^{-5}$ & $-3.84\underline{287728}
  \times 10^{-5}$\\[3pt]
&  $\omega_i\,\ell/\overline{U}\hspace{-.5mm}_e$ & $ +  3.014\underline{3827834}~
  \times 10^{-6}$ & $+2.335\underline{23}
  \times 10^{-6}$ & $-3.0\underline{7160739}
  \times 10^{-7}$\\[3pt]
\multirow{4}{*}{$\begin{array}{r}
\text{Relative} \\ \text{$|\omega|$-errors} \end{array} \left| \begin{array}{c}
{}\\{}\\{}\\{}\\{}
\end{array} \right.$ \hspace{-5mm}} &   $\epsilon_{L}$  & $ ~~~~~~~~~~+2.8\times 10^{-5~}$ & $ ~~~~ -1.7\times 10^{-4}$ & $ ~~~~~~~+4.3\times 10^{-5~}$\\[3pt]
&    $\epsilon_{N_x}$  & $ ~~~~~~~~~~+4.9\times 10^{-12}$ & $ ~~~~ +2.5\times 10^{-7}$ & $ ~~~~~~~-2.2\times 10^{-10}$ \\[3pt]
&     $\epsilon_{\mathrm{BC}}$ & $ ~~~~~~~~~~+1.5\times 10^{-10}$ & $ ~~~~ -3.6\times 10^{-7}$ & $ ~~~~~~~-6.9\times 10^{-11}$ \\[3pt]
&      $\epsilon_{y_{\mathrm{max}}}$ & $ ~~~~~~~~~~-2.3\times 10^{-4~}$  & $ ~~~~ -2.2\times 10^{-4}$ & $ ~~~~~~~-3.5 \times 10^{-4~}$
  \end{tabular}
  \end{ruledtabular}
  \caption{Mode properties and relative errors in the eigenvalue magnitude for the reference parameters given in Table \ref{tab:refparms} with respect to the parameter changes: $x_{\mathrm{out}}/\ell = 7.0 \times 10^5$ (fixing the density $N_x/L$); $N_x = 260$; the use of Dirichlet in-/outflow boundary conditions; and $y_{\mathrm{max}}/\ell = 1.4 \times 10^5$ (fixing $N_y=50$ and $y_i=4.0 \times 10^3$). The reported digits are truncated (not rounded) and those that are tainted by the largest reported error are underlined.}\label{tab:conv}
\vspace{-0mm}
  \end{center}
\end{table*}

The error introduced by the finite domain length, $\epsilon_L$, representing a primary source of error in the literature mentioned in the introduction, can be made an order of magnitude smaller than $\epsilon_{y_{\mathrm{max}}}$. Altering the resolution in the streamwise direction yields a very small error, $\epsilon_{N_x}$, due to the use of the spectral scheme with $N_x=260$ to $300$ nodes. It should be emphasised that these amounts of nodes are not at all necessary to obtain converging solutions for the reference case; using $N_x = 100$ nodes for mode 1 results in an error comparable to $\epsilon_{y_{\mathrm{max}}}$. The truncation boundary conditions represent the other primary uncertainty throughout the literature. By changing from Neumann to Dirichlet conditions, a remarkably small $\epsilon_{\mathrm{BC}}$ is obtained, that is equivalent to $\epsilon_{N_x}$. 

These results conclusively demonstrate that the obtained solutions are independent of the numerical set-up. The negligible influence of the streamwise domain length and truncation boundary conditions is observed to be directly related to the small amplitude of the eigenfunctions at the truncation boundaries. The spatial decay of the eigenfunctions within the domain allows placing the truncation boundaries at a \textit{far enough, but finite} distance, so that the eigeninformation is virtually unaffected. 

Only the modes along the main branch are found to converge; the side and downward branches persistently depend on the domain length. For increasing $L$, the side branch moves downward and modes that are originally positioned within the downward branch either merge with the main branch (and do converge thereafter) or they merge with the side branch. The fact that the side and downward branches do not converge is unexpected, because the corresponding eigenfunctions do decay toward all truncation boundaries, e.g.\ see modes 7 and 9 in Figs.~\ref{fig:existentialgraph}($f$,$g$). This illustrates that an eigenfunction's spatial decay toward the truncation boundaries is not a sufficient condition for the eigensolution to be independent of the numerical set-up.

\subsection{Dependence on $c_{\mathrm{g}}$}
\label{sec:variationcg}
\begin{figure*}[h]
\includegraphics[width=\textwidth]{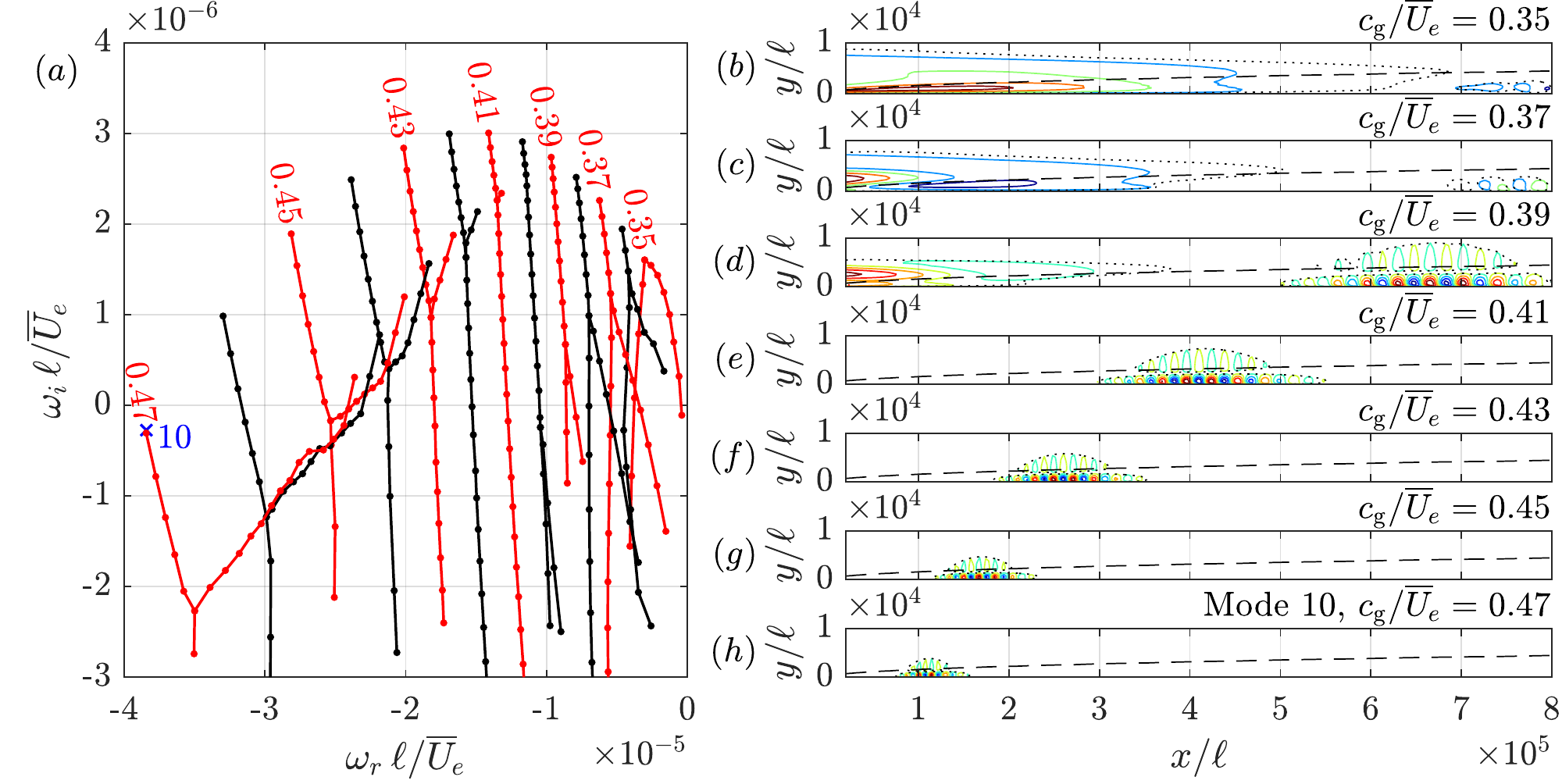}
\vspace{-0mm}
\caption{($a$) Relevant $\omega$-spectrum part and ($b$--$h$) isocontours of the real part of $\tilde{u}$ (coloured lines: from min- (blue) to maximum (red) with $\Delta = 2/9$, scaling the maximum to unity) and $|\tilde{u}|$ (black dotted, level: $1/9$) corresponding to the maximum $\omega_i$-eigenvalues along the labelled branches in ($a$) for indicative values of $c_{\mathrm{g}}$ ($\Delta c_{\mathrm{g}}/\overline{U}\hspace{-.5mm}_e = 0.01$), $\delta_{99}$-isocontour (black dashed).}
\label{fig:cgsweepgraphreal}
\vspace{-0mm}
\end{figure*}

Next, the movement of the converged part of the spectrum is studied while varying $c_\mathrm{g}$, see Fig.~\ref{fig:cgsweepgraphreal}($a$). The Doppler effect dictates that the frequency $\omega_r$ should decrease while $c_{\mathrm{g}}$ increases. 
System \eqref{eq:BiG} has real coefficients, rendering the spectrum symmetric about the $\omega_i$-axis.
The branch with $\omega_r<0$ moves as expected from the Doppler effect and is therefore considered. The reference case $c_{\mathrm{g}}/\overline{U}\hspace{-.5mm}_e = 0.415$ yielded the largest $\omega_i$-value. 

By increasing $c_{\mathrm{g}}/\overline{U}\hspace{-.5mm}_e \geq 0.41$, the main and side branches increase in extent and the eigenfunctions move upstream, see Figs.~\ref{fig:cgsweepgraphreal}($e$--$h$). While moving upstream, the streamwise extent of the eigenfunctions decreases and so does the streamwise wavelength. This is consistent with the boundary layer becoming thinner.

By decreasing $c_{\mathrm{g}}/\overline{U}\hspace{-.5mm}_e < 0.41$, the side branch coalesces with the main branch and the downward branch splits in two. The eigenfunctions move downstream and reach the outflow boundary. When close enough, the functions suddenly `latch' onto the outflow boundary and, simultaneously, an artificial structure emerges from the \textit{in}flow boundary. The point where the downwards branch splits reaches the top of the branch as $c_{\mathrm{g}}/\overline{U}\hspace{-.5mm}_e$ approaches 0.35, which causes the spectrum to have an arc-branch shape, as so described by \citet{lesshafft2017artificial}. As the spectrum attains the arc-branch shape, the latching tail from the inlet reaches the downstream structure, overwhelming the solution throughout the entire domain; all dynamics are then dominated by the artificial truncation boundary conditions. Tests show that solutions displaying this feature are strongly dependent on the artificial boundary conditions, domain size and $x$-resolution. Although the process changes by which the downward branch splits and how the eigenfunctions undergo latching, deploying fourth-order finite differences in the streamwise direction also results in artificial structures that reach from the in- to the outflow boundary at $c_{\mathrm{g}}/\overline{U}\hspace{-.5mm}_e = 0.35$.  

The results of \citet[\S IV.B.2, for $c_{\mathrm{g}} = 0$]{AlizardRobinet} suggest that arc-shaped spectra obtained for too small $c_{\mathrm{g}}$ approach a continuum as the streamwise domain length tends to infinity. Numerous analyses presented in the literature are performed in the stationary reference frame and result in arc-shaped spectra. The present analysis suggests that the domain truncation has had a non-negligible artificial impact on these results. 

It is concluded that, keeping the reference domain length fixed, a large enough $c_{\mathrm{g}}$-value is required to prevent the eigenfunctions from reaching the outlet truncation boundary and eliminate the unwanted dependency on the numerical set-up. 

\subsection{The limit $c_\mathrm{g}\rightarrow 0$}
By increasing the domain length, the eigenfunctions can propagate further downstream and $c_\mathrm{g}$ could be further decreased, attempting to recover the stationary reference frame. 

\begin{figure*}[h]
\vspace{-0mm}
\centering
\includegraphics[width=0.95\textwidth]{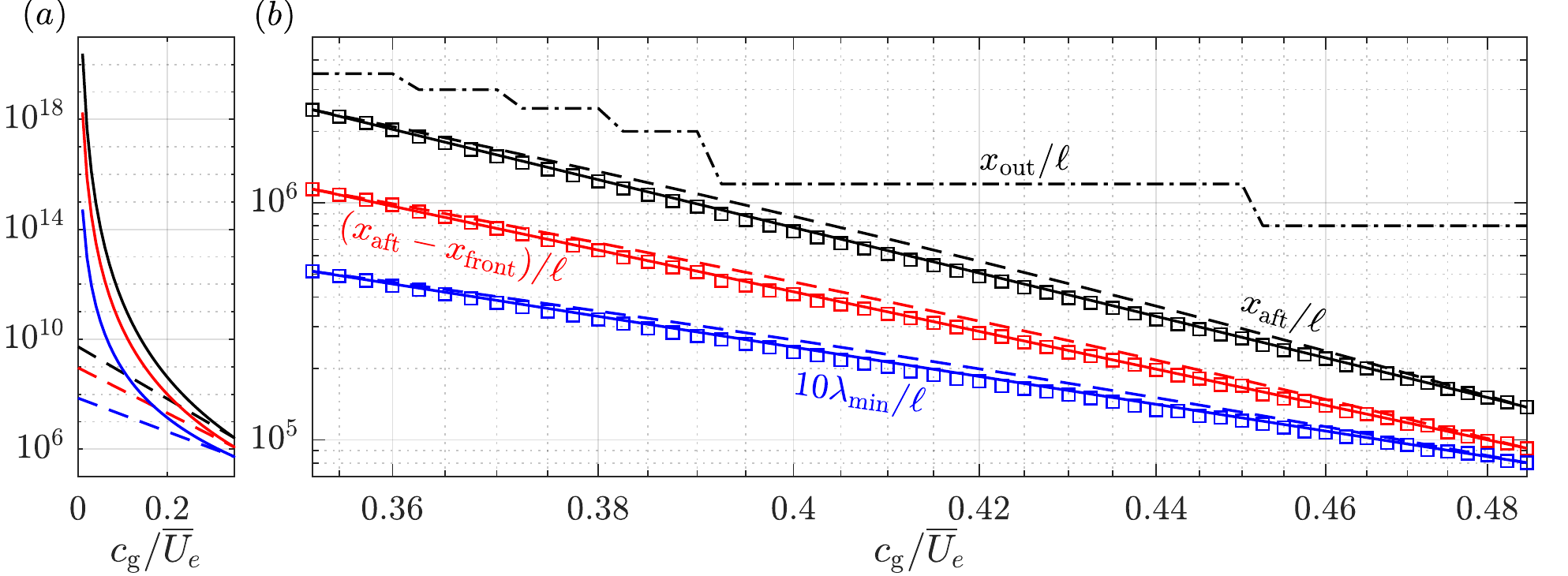}
\vspace{-0mm}
\caption{For the most unstable $\tilde{u}$-eigenfunction: aft-most location ($x_{\mathrm{aft}}$, black), streamwise extent ($x_{\mathrm{aft}} - x_{\mathrm{front}}$, red) and minimum wavelength ($10\times \lambda_\mathrm{min}$, blue) versus $c_\mathrm{g}$. Measured values (symbols), power (solid lines) and exponential (dashed) fits and outflow boundary $x_{\mathrm{out}}$ (dash-dotted). ($a$) Trend-extrapolation as $c_\mathrm{g}\rightarrow 0$ (minimum: $0.025\overline{U}\hspace{-.5mm}_e$) and ($b$) data and fits.}
\label{fig:variationWPchar2}
\vspace{-0mm}
\end{figure*}

`Latching' is from now on identified with the emergence of an artificial inlet structure. It is observed when the $\tilde{u}$-eigenfunction attains an $O(10^{-3})$ relative magnitude at the outflow boundary for the reference domain length. Hence, the most up- and downstream position of the wavepacket, $x_{\mathrm{front}}$ and $x_{\mathrm{aft}}$, are respectively defined to be the first and last positions where this level is measured. 
Furthermore, the minimum wavelength $\lambda_{\mathrm{min}}$ (represented by the real and imaginary part of $\tilde{u}$) for $x\in[x_{\mathrm{front}},x_{\mathrm{aft}}]$ is measured. The domain length is increased (fixing $N_x/L$) to properly capture the most unstable solution along the main branch. The measured variation of $x_{\mathrm{aft}}- x_{\mathrm{front}}$, $x_{\mathrm{aft}}$ and $\lambda_{\mathrm{min}}$ with $c_{\mathrm{g}}$ is shown in Fig.~\ref{fig:variationWPchar2}($b$); all increase with decreasing $c_{\mathrm{g}}$. Their growth rates are quantified by fitting a power and exponential law; the resulting parameters are reported in Table~\ref{tab:fits}. 

If the power-law trend holds in the limit, the quantities approach infinity when $c_{\mathrm{g}}\rightarrow 0$, as illustrated in Fig.~\ref{fig:variationWPchar2}($a$), rendering resolving localised wavepackets in the stationary reference frame impossible. The conjecture that this is impossible is supported by the theory presented by \citet[equation 8.10]{GrootPhDThesis}. Based on the results that could be obtained with the available computational resources, significantly larger correlations were found for the power versus exponential law, see Table~\ref{tab:fits}. Furthermore, independent of the fitting law used, $\lambda_{\mathrm{min}}$ increases at a much lower rate than $x_{\mathrm{aft}}-x_{\mathrm{front}}$. Therefore, both the $x$-resolution and the domain length have to be increased as $c_{\mathrm{g}}\rightarrow 0$, so that the problem becomes computationally intractable when $c_{\mathrm{g}}$ becomes small. An insufficient $x$-resolution was also observed to cause latching to the in-/outflow boundaries. 

\begin{table}[h]
  \begin{center}
\vspace{-0mm}
\def~{\hphantom{0}}
\begin{ruledtabular}
  \begin{tabular}{r c c c c c}
    & $x_{\mathrm{aft}}$ & & $x_{\mathrm{aft}} - x_{\mathrm{front}}$ & & $\lambda_\mathrm{min}$ \\[3pt]
Power law: & $-9.0$ (0.9999) & & $-7.9$ (0.9998) & & $-5.8$ (0.9995) \\
Exponential law: & $-9.5$ (0.9979) & & $-8.2$ (0.9981) & & $-6.1$ (0.9972)
  \end{tabular}
\end{ruledtabular}
\vspace{-0mm}
  \caption{Fit parameters based on the outermost $c_{\mathrm{g}}$ data points in Fig. \ref{fig:variationWPchar2} for $x_{\mathrm{aft}}$, $x_{\mathrm{aft}} - x_{\mathrm{front}}$ 
  and $\lambda_\mathrm{min}$ of the wavepacket: $p$ in $a c_{\mathrm{g}}^p$ for the power law, $\varepsilon$ in $b \mathrm{e}^{\varepsilon c_{\mathrm{g}}}$ for the exponential law, the Pearson correlation coefficients are given in brackets.} 
\vspace{-0mm}
  \label{tab:fits}
  \end{center}
\end{table}

\section{Link with local and non-local methods}
\label{sec:linkLSTPSE}
As opposed to the approach used in literature \citep{AlizardRobinet,Rodriguez2010}, this section demonstrates the link between local (LST), non-local (PSE) and global stability methods for Blasius flow without making use of the BiGlobal eigeninformation. The converged modes appear as \textit{discrete} (i.e.\ not \textit{forced continuum}) modes, which permits their approximation via the global mode theory developed by \citet{monkewitz1993global}. From a physical perspective, this can be justified as follows. A convective instability appears as an absolute instability in a moving reference frame \citep{HuerrePMonkewitz1985,SchmidHenningson}. In turn, the existence of an absolute instability is a necessary condition for the existence of a global instability \citep{HuerreMonkewitz}. To demonstrate these instability natures, a spatio-temporal stability framework must be used. For conciseness, this framework will here be recited in recipe form, after describing the used (non-)local stability approaches, see \citet{monkewitz1993global} for all details.

The LST and PSE problems are discretised as consistently as possible with respect to the BiGlobal problem. 
For PSE, the stabilised discretisation method proposed by \citet{andersson1998stabilization} is used.
The streamwise wavenumber $\sigma$ equals:
\begin{equation}
\begin{array}{rl}
\text{LST:}& \quad \sigma = \alpha;\\
\text{PSE:}& \quad \sigma =  \alpha + \alpha_{\mathrm{aux}},
\end{array} \quad  \text{where:}\quad \alpha_{\mathrm{aux}} = -\mathrm{i} \int_0^{y_{\mathrm{max}}} \tilde{u}^* \frac{\partial \tilde{u}}{\partial x}\,\mathrm{d}y\Bigg/\int_0^{y_{\mathrm{max}}} |\tilde{u}|^2\,\mathrm{d}y,
\end{equation}

\noindent and $\alpha$ is the streamwise wavenumber in the standard perturbation ansatzes \citep{MackAGARD,Herbert1997} and the star denotes complex conjugation. For PSE, the growth in the shape function is accounted for with $\alpha_{\mathrm{aux}}$, which is minimised up to $O(10^{-10})$ relative errors. Both problems are solved for a frequency $\bar{\omega}$ and location $\bar{x}$ corresponding to the stationary frame of reference, such that $\sigma = \sigma(\bar{x},\bar{\omega})$. The frequency in the moving frame of reference is obtained through the Doppler shift formula: $\omega(\bar{x},\bar{\omega}) = \bar{\omega} - \sigma(\bar{x},\bar{\omega}) c_{\mathrm{g}}$, equivalent to equation \eqref{eq:generalisedDoppler}. In the moving reference frame, the solutions have a non-convective nature. Therefore, the PSE problem would diverge if solved in that reference frame; resorting to the stationary reference frame allows circumventing this issue entirely.

The global frequency is obtained by manipulating the LST/PSE solutions as follows:
\begin{enumerate}
\item Find $\bar{\omega}\in \mathbb{C}$ for which $|\mathrm{d}\omega/\mathrm{d}\sigma| = 0$, while fixing $\bar{x}$. This is equivalent to $|\mathrm{d} \omega/\mathrm{d} \bar{\omega}| = 0$, because ${\mathrm{d} \omega}/{\mathrm{d} \sigma} = c_{\mathrm{g}}/((\mathrm{d} \bar{\omega}/\mathrm{d} \omega) - 1)$. The corresponding solutions are indicated by the subscript $0$; $\omega=\omega_0$ represents a saddle point when $\omega$ is graphed versus $\sigma$ and $\sigma_0$ is a double root of the dispersion relation. Note that $\omega_0$ and $\sigma_0$ are both a function of $\bar{x}$. 
\item Find $\bar{x}$ for which $|\mathrm{d}\omega_0/\mathrm{d}\bar{x}| = 0$. For this $\bar{x}$-value, $\omega_0(\bar{x})$ displays a cusp in the $\omega$-plane and this cusp-point, denoted by $\omega_\mathrm{g}$, approximates one BiGlobal eigenvalue. 
\end{enumerate}

\begin{figure*}[h]
\centering
\includegraphics[width=\textwidth]{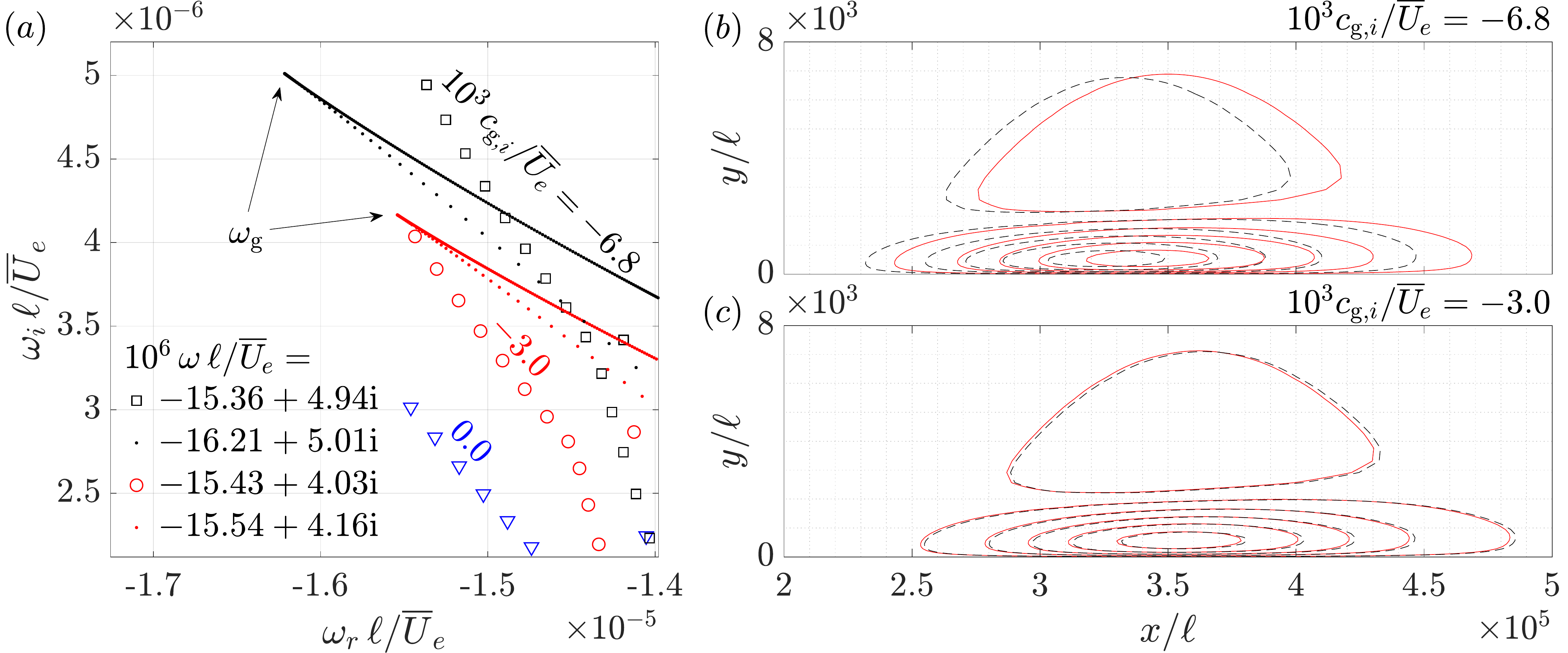}
\vspace{-0mm}
\caption{($a$) Comparison of the BiGlobal spectra (open symbols) for the complex $c_{\mathrm{g}}$-values for which cusp-branches could be obtained with PSE (red dots) and LST (black dots); equal colours indicate equal $c_{\mathrm{g},i}$-values, $\omega$-values for LST and PSE cusps and most unstable BiGlobal eigenvalues are presented in the legend (for $c_{\mathrm{g},i} = 0$: $\omega\,\ell/\overline{U}\hspace{-.5mm}_e = (-15.46 + 3.01\mathrm{i})\times 10^{-6}$). ($b$,$c$) Comparison of the reconstruction of the BiGlobal $|\tilde{u}|$-eigenfunction (red solid, levels: 1, 3, $\ldots$ 9, scaling the maximum to 10) with LST ($b$, black dashed) and PSE ($c$, black dashed) for the respective $c_{\mathrm{g},i}$-values used in ($a$). The value $c_{\mathrm{g},r}/\overline{U}\hspace{-.5mm}_e = 0.415$ is used for all presented cases.} 
\label{fig:comparisonLSTPSE}
\vspace{-0mm}
\end{figure*}

The criteria (i) and (ii) were checked numerically by evaluating the solutions for increasingly denser $\bar{\omega}$- and $\bar{x}$-sequences, respectively. The criteria were deemed satisfied if the magnitude of the derivatives $|\mathrm{d}\omega/\mathrm{d}\bar{\omega}|$ and $|\mathrm{d}\omega_0/\mathrm{d}\bar{x}|$ was of $O(10^{-6})$ and $O(10^{-14}\overline{U}\hspace{-.5mm}_e/\ell^2)$
, respectively. No $\omega_{\mathrm{g}}$-cusp could be found for the real-valued $c_{\mathrm{g}} = 0.415\overline{U}\hspace{-.5mm}_e$. Therefore $c_{\mathrm{g},r}$ was fixed (to the reference value 0.415$\overline{U}\hspace{-.5mm}_e$) and non-zero $c_{\mathrm{g},i}$-values were permitted. This is equivalent to the use of a complex spatial $\bar{x}$-coordinate by \citet{monkewitz1993global}. This did yield LST and PSE solutions satisfying both criteria ($10^{3}c_{\mathrm{g},i}/\overline{U}\hspace{-.5mm}_e = -3.0$ for PSE and $-6.8$ for LST). Note that the use of complex group speeds is necessary only to perform the comparison with LST and PSE; having retrieved converged BiGlobal modes for real $c_{\mathrm{g}}$ demonstrates that complex group speeds can be circumvented completely through the use of the moving reference frame. An interpretation of the complex group speeds and complex spatial coordinates is therefore immaterial. Fig.~\ref{fig:comparisonLSTPSE}($a$) displays the cusped $\omega_0(\bar{x})$-branches and the BiGlobal spectra for the corresponding complex $c_{\mathrm{g}}$-values. 

Comparing the $\omega_{\mathrm{g}}$-values to the BiGlobal eigenvalues (given in the legend), the minimum distance is smaller for PSE (red) than for LST (black). Similarly, a smaller $c_{\mathrm{g},i}$-value is required for PSE than LST. Both the smaller distance of the $\omega_{\mathrm{g}}$-value and smaller $c_{\mathrm{g},i}$-value are argued to be caused by the smaller model error in the PSE over the LST approach, i.e.\ due to non-parallel effects. The reconstruction of the BiGlobal eigenfunction with the (non-)local solutions is shown in Figs.~\ref{fig:comparisonLSTPSE}($b$,$c$), obtained by fixing the frequency in the moving reference frame to $\omega_{\mathrm{g}}$ and selecting the $\alpha$-branches that represent decaying solutions in the up- and downstream directions \citep[equation 3.16]{monkewitz1993global}. The obtained $|\tilde{u}|$-functions closely resemble the BiGlobal equivalents. The structure found with LST lies significantly upstream of the BiGlobal one, while a striking match is obtained with PSE. These comparisons reflect the found differences in the $\omega_{\mathrm{g}}$- and $c_{\mathrm{g},i}$-values and they confirm that the most unstable BiGlobal modes are recovered.

These results establish the link between the BiGlobal and $\text{(non-)}$local stability approaches for convective mechanisms, specifically. It moreover demonstrates that, while Tollmien-Schlichting waves are a convective instability in the stationary reference frame, they represent a global instability mechanism in a moving reference frame.

In conclusion, the BiGlobal stability method formulated in the moving reference frame does exactly what it is supposed to do: it localises global instability mechanisms without having to go through the complicated $\sigma$-saddle- and $\omega$-cusp-point-finding algorithms and without having to deal with esoteric complex group speeds or complex spatial coordinates.

\section{Conclusion}
\label{sec:conclusion}
By solving the BiGlobal problem in a moving frame of reference, we obtain eigensolutions that converge numerically for a sufficiently large, but finite resolution and domain length. These solutions appear as discrete modes in the eigenvalue spectrum, which enabled us to independently approximate the BiGlobal eigenvalues with $\text{(non-)}$local stability methods, i.e.\ without using the BiGlobal results as input. Moreover, we demonstrate that retrieving converged eigensolutions in the stationary reference frame is likely impossible for the examined base flow case. 

A moving reference frame renders developing base flows unsteady. While currently not accounting for the related effects, the present methodology and results establish a reliable point of departure for their quantification by performing time-integration in the future. Further investigations should be focused on more closely establishing the link between the absolute and global instability characteristics and reproducing the equivalents of the neutral and amplification ($N$-factor) curves.

\section*{Acknowledgements}
The authors acknowledge the funding provided by the FNRS-FRIA fellowship granted to Sébastien E.M. Niessen and thank Henk Schuttelaars and Stefan Hickel at Delft University of Technology, Fabio Pinna at the Von K\'arm\'an Institute for Fluid Dynamics, Vincent Terrapon at Universit\'e de Li\`ege and Ethan Beyak, Andrew Riha and Helen Reed at Texas A\&M University for the useful discussions.

\bibliography{PRLsBiGp}
\end{document}